%% file: ms.tex
\documentclass[conference,letterpaper]{IEEEtran}

\usepackage[utf8]{inputenc} 
\usepackage[T1]{fontenc}
\usepackage{url}
\usepackage{ifthen}
\usepackage{cite}
\usepackage[cmex10]{amsmath} 
\usepackage{comment}
\usepackage[normalem]{ulem}

\input{preamble}

\usetikzlibrary{arrows}
\tikzset{
     invisible/.style={opacity=0},
     visible on/.style={alt={#1{}{invisible}}},
     alt/.code args={<#1>#2#3}{%
       \alt<#1>{\pgfkeysalso{#2}}{\pgfkeysalso{#3}} %
     },
   }

\usepackage{ marvosym }
\usepackage{bm}

\newcommand{\pbar}{\ensuremath{\bar{p}}}
\newcommand{\qbar}{\ensuremath{\bar{q}}}
\newcommand{\na}{\ensuremath{n^\alpha}}
\newcommand{\nb}{\ensuremath{n^\beta}}
\newcommand{\Ra}{R_\alpha}

\newcommand{\Eoha}{\ensuremath{\hat{E}_0^\alpha}}
\newcommand{\Egha}{\ensuremath{\hat{E}_G^\alpha}}
\newcommand{\Eohan}{\ensuremath{\hat{E}_r^{\alpha,n}}}
\newcommand{\Eorha}{\ensuremath{\hat{E}_r^\alpha}}
\newcommand{\Erha}{\ensuremath{\hat{E}_R^\alpha}}

\newcommand{\Pxn}{\ensuremath{P_{X,n}}}
\newcommand{\Pyn}{\ensuremath{P_{Y,n}}}
\newcommand{\Pyx}{\ensuremath{P_{Y\given X}}}

\newcommand{\Pxdn}{\ensuremath{P_{X_o,n}}}

\newcommand{\Pxcw}{\ensuremath{P_{X^n}}}

\newcommand{\Pycwxcw}{\ensuremath{P_{Y^n\given X^n}}}

\newcommand{\Pzn}{\ensuremath{P_{Z,n}}}
\newcommand{\Pzx}{\ensuremath{P_{Z\given X}}}
\newcommand{\Pzd}{P_{Z_o}}
\newcommand{\Pzdn}{\ensuremath{P_{Z_o,n}}}
\newcommand{\Pzcw}{\ensuremath{P_{Z^n}}}

\newcommand{\Pzprod}{\ensuremath{P^n_{Z}}}
\newcommand{\Pzdprod}{\ensuremath{P^n_{Z_o}}}
\newcommand{\PzcwtcC}{\ensuremath{P_{Z^n\given \tcC}}}

\newcommand{\tcC}{\tilde{\cC}}

\newcommand{\muvec}{\ensuremath{\bm{\mu}}}

\usepackage[printonlyused,nolist]{acronym}
\begin{acronym}
    \acro{AWGN}{additive white Gaussian noise}
    \acro{LLR}{log-likelihood ratio}
    \acro{WLLN}{weak law of large numbers}
    \acro{DSP}{digital signal processing}
    \acro{BRGC}{binary reflected Gray code}
    \acro{PDM}{polarization division multiplexing}
    \acro{PMF}{probability mass function}
    \acro{RV}{random variable}
    \acro{PDF}{probability density function}
    \acro{BSC}{binary symmetric channel}
    \acro{DMS}{discrete memoryless source}
    \acro{DMC}{discrete memoryless channel}
    \acro{KL}{Kullback-Leibler}
    \acro{ML}{maximum-likelihood}
    \acro{RHS}{right-hand side}
    \acro{LHS}{left-hand side}
    \acro{VP}{vanishing power}
\end{acronym}

\begin{document}
\title{Stealth Communication with Vanishing Power\\ over Binary Symmetric Channels}

 \author{%
   \IEEEauthorblockN{Diego Lentner and Gerhard Kramer}
   \IEEEauthorblockA{Institute for Communications Engineering, 
                     Technical University of Munich, 80290 Munich, Germany \\
                     Email: \{diego.lentner, gerhard.kramer\}@tum.de}
 }
 
\maketitle

\begin{abstract}
A framework for stealth communication with vanishing power (VP) is presented by studying binary symmetric channels. Coding theorems are proved by modifying Gallager's error exponents for VP and by applying resolvability exponents. The analysis unifies and generalizes existing rate bounds for covert and stealth communication.
\end{abstract}

\section{Introduction}
Covert communication \cite{bash2013limits} refers to a scenario where a sender Alice communicates with a receiver Bob without a third party, Warren, being able to detect the communication. In contrast to the secrecy problem, it is not the content of the message that Alice and Bob want to hide from Warren but the presence of the message itself. Bash \emph{et al.} \cite{bash2013limits} showed that on the order of $\sqrt{n}$ bits can be covertly communicated in $n$ channel uses over \ac{AWGN} channels. This \emph{square root law} also applies to \acp{DMC}
\cite{wang2016fundamental,bloch2016covert}.

Covertness can be measured by the informational divergence of two types of channel output statistics: those when a meaningful message is transmitted and those when a sequence of ``zero'' symbols is transmitted, where the ``zero'' symbol usually represents the absence of energy. Second-order asymptotics for various covertness measures are derived in \cite{tahmasbi2019first}. Covert communication may require shared randomness between Alice and Bob in the form of a secret key, unless Warren's channel from Alice is noisier than Bob's, as shown in \cite{che2013reliable}, \cite{che2014reliable} for \acp{BSC}.

\emph{Stealth communication} generalizes covert communication by discarding the requirement that Alice must be silent when not communicating information to Bob, i.e., Alice is free to transmit symbols other than the ``zero'' symbol. The idea is that Alice confuses Warren by sending obfuscating symbols. Obfuscation is an old technique to enhance \emph{privacy}, e.g., it can hide personal information such as mobility patterns or web browsing behavior.

One can show formally~\cite{hou2014effective,Hou-Kramer-Bloch-2017} that obfuscation can break the square root law, and in fact communication with positive rate is possible without Warren being able to detect meaningful communication. The price that Alice pays is that she must consume more energy than for covert communication, and we thus arrive at a capacity-cost tradeoff. This tradeoff depends on which obfuscation patterns are permitted, and we will consider obfuscation strings consisting of independent and identically distributed (i.i.d.) channel symbols.

The main contribution of this work is two-fold.
\begin{enumerate}
\item We introduce a framework for stealth communication that includes previously treated scenarios as special cases. In particular, we are interested in using vanishing power, as for covert communication, but with energy that scales as $n^\alpha$, $0\le\alpha<1$, with blocklength $n$. Observe that covert communication has
$\alpha\le1/2$ while stealth communication as treated in~\cite{hou2014effective,Hou-Kramer-Bloch-2017} has $\alpha=1$.
\item We prove coding theorems by using suitably modified Gallager exponents. This gives an alternative, and we believe simpler, approach to prove and understand achievability
as compared to previous work.
\end{enumerate}

This paper is organized as follows. Sec.~\ref{sec:section2} introduces notation and classic error exponents. In Sec.~\ref{sec:section3}, we derive achievable codebook scaling constants for \ac{VP} communication by using modified error exponents. We apply these results in Sec.~\ref{sec:section4} to prove achievability of VP stealth communication. Finally, we compare our results for the covert communication case with bounds from \cite{wang2016fundamental,bloch2016covert,che2014reliable} in Sec.~\ref{sec:conclusion}.

\section{Preliminaries} \label{sec:section2}
\subsection{Notation}
Random variables are denoted by upper case letters and their realizations by the corresponding lower case letters. Finite sequences of random variables are written with a superscript indicating the number of symbols of the sequence, e.g., $X^n = X_1,\ldots,X_n$. Let $X$ be a discrete random variable with probability distribution $P_X$ and alphabet $\cX$. If the symbols $X_i$, $i=1,\ldots,n$, are i.i.d. according to $P_X$, then the distribution of $X^n$ is $P_{X^n}(x^n) = \prod_{i=1}^nP_X(x_i) \equivalent P_X^n(x^n)$. 
For any two probability distributions $P_X$ and $P_{\tilde{X}}$ on $\cX$ where $P_X\ll P_{\tilde{X}}$, i.e. $P_{\tilde{X}}(x)=0 \Rightarrow P_X(x)=0$ for any $x\in\cX$, the informational divergence between $P_X$ and $P_{\tilde{X}}$ is
\begin{align}
    \id{P_X}{P_{\tilde{X}}} \equivalent \sum_{x: P_X(x)>0}P_X(x)\log\frac{P_X(x)}{P_{\tilde{X}}(x)}
\end{align}
and  their variational distance is defined as 
\begin{equation}
\vd{P_X}{P_{\tilde{X}}} \equivalent \frac{1}{2}\sum_{x\in\cX}\abs{P_X(x) - P_{\tilde{X}}(x)} . \label{eq:vd_definition}
\end{equation}
Informational divergence and variational distance are related by Pinsker's inequality:
\begin{equation}
\vd{P_X}{P_{\tilde{X}}}^2 \leq \frac{1}{2} \id{P_X}{P_{\tilde{X}}} \label{eq:pinsker_ineq} \,.
\end{equation}
The chi-squared distance of $P_X$ and $P_{\tilde{X}}$ is
\begin{align}
\chsq{P_X}{P_{\tilde{X}}} \equivalent \sum_{x\in\cX} \frac{\left(P_X(x)-P_{\tilde{X}}(x)\right)^2}{P_{\tilde{X}}(x)} .
\end{align}
The mutual information of $X$ and $Y$ is denoted
$I(P_X;\Pyx)$.

\subsection{Error Exponents}\label{sec:error_exponent}
Let $W$ be Alice's message and let $\hat{W}$ be Bob's estimate of this message. Gallager used a random coding argument to show that if each message $W$ selects a codeword from a code of cardinality $M$ and length $n$, then the worst-case error probability under \ac{ML} decoding over a noisy channel $\Pycwxcw$ can be bounded as \cite[Ch.~5]{gallager1968information}
\begin{align}
   &\Pr\left[\hat{W}\ne 1 \given W=1\right] \nonumber\\
    &\le (M-1)^\rho \sum_{y^n} \left\{ \sum_{x^n} \Pxcw(x^n) \Pycwxcw(y^n \given x^n)^{\frac{1}{1+\rho}} \right\}^{1+\rho}
    \label{eq:rcb_general}
\end{align}
where $\rho$, $0\leq\rho\leq 1$, is an optimization parameter.

Consider a \ac{DMS} with probability distribution $P_X$ and a \ac{DMC} $\Pyx$. Define the code rate as $R = \frac{\log M}{n}$. We have $(M-1)^\rho \le M^\rho = e^{n R \rho}$ and \eqref{eq:rcb_general} becomes (see \cite{gallager1968information})
\begin{align}
   \Pr\left[\hat{W}\ne 1 \,|\, W=1\right]& \le e^{-n E_G(R,P_X)}
\end{align}
where the error exponent $E_G(R,P_X)$ is defined as
\begin{align}
   & E_G(R,P_X) = \max_{0\le\rho\le1} \left[ E_0(\rho,P_X) - \rho R \right]
   \label{eq:Eg_definition}\\
   & E_0(\rho,P_X) = -\log \sum_{y} \left\{ \sum_{x}
   P_X(x) \Pyx(y | x)^{\frac{1}{1+\rho}} \right\}^{1+\rho}.
   \label{eq:E0_definition}
\end{align}

\section{Error Exponents for VP Communication} \label{sec:section3}
This section considers classic point-to-point communication over a BSC and with \ac{VP}. Let BSC$(p)$ denote a \ac{BSC} with cross-over probability $p$. We define the energy of the binary sequence $X^n$ as its Hamming weight $\sum_{i=1}^{n}X_i^2$. In the following, the code length $n$ is a free parameter and we transmit only one codeword (one-shot analysis).

\subsection{Information Rate Analysis} \label{sec:finite_energy_tx_model}
Suppose we have the average block power constraint
\begin{align}
    \frac{1}{n}\expop\left[\sum_{i=1}^{n}X_i^2\right] \le \frac{a\na}{n}
    \label{eq:power-constraint}
\end{align}
where $0 \leq \alpha \leq 1$ and $0<a$. The constraint can be satisfied by choosing the channel input distribution as $\Pxn(1) = 1-\Pxn(0) = \frac{a\na}{n}$. Note that the distribution $\Pxn$ directly depends on the choice of the blocklength $n$, which we emphasize with the additional subscript. Clearly, if $0\le\alpha<1$ then the power of $X^n$ will vanish for $n\to\infty$. We therefore refer to signaling with $0\le\alpha<1$ as \ac{VP} transmission.

We assess how much information can be transmitted with \ac{VP} over a BSC$(p)$. Let the transmitted signal $X^n$ be distributed according to $\Pxn$. The receiver observes the binary sequence $Y^n$ which is distributed as
\begin{align}
    \Pyn(0) &= \left(1-\frac{a\na}{n}\right)\pbar + \frac{a\na}{n} p  \nonumber \\
    &= \pbar - (1-2p)\cdot \frac{a\na}{n}\\
    \Pyn(1) &= \left(1-\frac{a\na}{n}\right) p + \frac{a\na}{n} \pbar \nonumber \\
    &= p + (1-2p)\cdot \frac{a\na}{n}
\end{align}
where we introduced the shorthand $\pbar = 1-p$. 
The mutual information is
\begin{align}
    & I(\Pxn;\Pyx) = H(\Pyn) - H_2(p) \nonumber \\
    &\qquad = H_2\left(p + (1-2p)\cdot \frac{a\na}{n}\right) - H_2(p) \nonumber \\
    &\qquad\approx (1-2p) \frac{a\na}{n} \cdot \log\frac{\pbar}{p} \label{eq:mi_alpha_simplified}
\end{align}
where $H_2(p)=-p\log p - (1-p)\log(1-p)$ and where we have used the first-order Taylor expansion
\begin{equation}
    H_2(x)\rvert_{x=p} \approx H_2(p) + (x-p) \cdot \left.\frac{\partial H_2(x)}{\partial x}\right\vert_{x=p}.
\end{equation}

\subsection{Modified Random Coding Exponent} \label{sec:mod_error_exponents}

Directly applying the error exponent framework introduced in Sec.~\ref{sec:error_exponent} to our model has $E_0(\rho,\Pxn)$ scaling with $\na/n$ which goes to zero as $n$ increases. To get a more meaningful exponent, we normalize \eqref{eq:E0_definition} by the scaling factor $\na/n$ and compute
\begin{align}
\Eoha(\rho,\Pxn) &= \lim_{n\to\infty} \frac{n}{\na} E_0(\rho,\Pxn) \,.
\label{eq:E0ha_definition}
\end{align}
Now define $\Ra=\frac{1}{\na} \log M$ and a \emph{modified} error exponent
\begin{equation}
\Egha(\Ra,\Pxn) = \max_{0\leq\rho\leq 1} \left(\Eoha(\rho,\Pxn) - \rho\Ra\right)
\label{eq:Egha_definition}
\end{equation}
to describe the error probability decay with $\na$ as
\begin{equation}
\Pr\left[\hat{W}\ne 1 \,|\, W=1\right] \leq e^{-\na\Egha(\Ra,\Pxn)}
\end{equation}
for large $n$.

In the following we show that the modified error exponents exhibit similar properties as the well-studied Gallager exponents reviewed in Sec.~\ref{sec:error_exponent}. For the \ac{BSC} channel model with input $\Pxn$ as defined in Sec.~\ref{sec:finite_energy_tx_model}, the expression \eqref{eq:E0ha_definition} can be explicitly derived:
\begin{align}
    &\Eoha(\rho,\Pxn) \nonumber = \lim_{n\to\infty} \tfrac{n}{\na} \left[-\log\left(\left(\left(1-\tfrac{a\na}{n}\right)\pbar^{\frac{1}{1+\rho}}+\tfrac{a\na}{n}p^{\frac{1}{1+\rho}}\right)^{1+\rho} \right.\right. \nonumber\\
    &\qquad  +\left.\left. \left(\left(1-\tfrac{a\na}{n}\right)p^{\frac{1}{1+\rho}}+\tfrac{a\na}{n}\pbar^{\frac{1}{1+\rho}}\right)^{1+\rho}\right)\right] \label{eq:E0ha_simplified_deriv1}\\
    &\quad = \begin{cases}(1+\rho) a \left(\pbar^{\frac{1}{1+\rho}} - p^{\frac{1}{1+\rho}}\right) \left(\pbar^{\frac{\rho}{1+\rho}} - p^{\frac{\rho}{1+\rho}}\right),& \alpha< 1\\
E_0(\rho,\Pxn) ,& \alpha= 1 \,.\end{cases}
    \label{eq:E0ha_simplified}
\end{align}

The complete derivation of \eqref{eq:E0ha_simplified} involves L'Hospital's rule
and is omitted due to space limitations.
For the extremal values of $\rho$, we have for $0\leq\alpha<1$:
\begin{align}
\lim_{\rho\to 1} \Eoha(\rho,\Pxn) &= 2a(\sqrt{\pbar}-\sqrt{p})^2 \\
\lim_{\rho\to 0} \Eoha(\rho,\Pxn) &= 0 \,.
\end{align}

Finally, the maximum scaling constant $\Ra$ for which the modified error exponent \eqref{eq:Egha_definition} is positive, and therefore the error probability vanishes for large $n$, is given by
\begin{align}
    &R_{\alpha,\text{max}}(\Pxn) = \left.\frac{\partial \Eoha(\rho,\Pxn)}{\partial\rho}\right\vert_{\rho=0} \nonumber\\ 
    &= a \left\{ \left(\pbar^{\frac{1}{1+\rho}}-p^{\frac{1}{1+\rho}}\right) \left(\pbar^{\frac{\rho}{1+\rho}}-p^{\frac{\rho}{1+\rho}}\right) + \frac{1}{1+\rho}\cdot \right.\nonumber\\
    &\qquad \left[\left(\pbar^{\frac{1}{1+\rho}}-p^{\frac{1}{1+\rho}}\right)\left(\pbar^{\frac{\rho}{1+\rho}}\cdot\log\pbar-p^{\frac{\rho}{1+\rho}}\cdot\log p\right) \right.\nonumber\\
    &\qquad \left.\left. - \left(\pbar^{\frac{1}{1+\rho}}\cdot\log\pbar-p^{\frac{1}{1+\rho}}\cdot\log p\right)\left(\pbar^{\frac{\rho}{1+\rho}}-p^{\frac{\rho}{1+\rho}}\right) \right] \right\}_{\rho=0} \nonumber\\
    &= a(1-2p)\log\frac{\pbar}{p} \label{eq:Ra_max} .
\end{align}
Observe that the \ac{RHS} of \eqref{eq:Ra_max} is the same as the \ac{RHS} of \eqref{eq:mi_alpha_simplified} after normalizing by $\na/n$. The error probability thus decays exponentially with $\na$ if
\begin{align}
\na\Ra \lessapprox n I(\Pxn;\Pyx)
\end{align}
which for $\alpha=1$ reduces to $R<I(P_X;\Pyx)$.

\section{Stealth Communication with VP Obfuscation} \label{sec:section4}
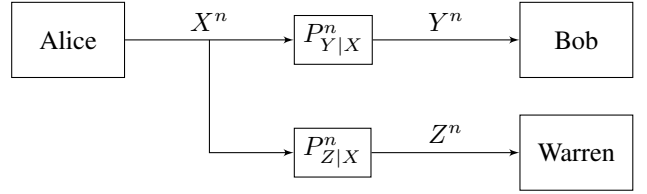
\begin{figure}
    \centering
 	\input{./fig1.tex}
    \caption{Stealth communication problem.} \label{fig:stealth_problem}
\end{figure}

Consider now the stealth communication problem depicted in Fig.~\ref{fig:stealth_problem}. Alice wants to transmit a message reliably to Bob over the memoryless channel $\Pyx$ without being detected by Warren. Warren observes the output of the channel $\Pzx$ and makes a binary hypothesis test whether Alice has transmitted information or obfuscation symbols. Suppose the channel $\Pyx$ from Alice to Bob is a BSC$(p)$, and the channel $\Pzx$ from Alice to Warren is a BSC$(q)$. We do not restrict the values of $p$ and $q$ other than $p\le1/2$ and $q\le1/2$. Suppose that Alice sends with \ac{VP} as in \eqref{eq:power-constraint}
when transmitting information, and that she sends i.i.d. sequences $X^n$ with \ac{VP} given by 
\begin{equation}
\frac{1}{n}\expop\left[\sum_{i=1}^{n}X_i^2\right] = \frac{b\nb}{n}, \quad 0 \leq \beta < 1, \, 0<b \,. \label{eq:power-constraint-o}
\end{equation}
when transmitting obfuscation symbols.
In the following, we derive conditions on $a,\alpha,b,\beta$ to achieve stealth communication for both uncoded and coded transmission.

\subsection{Uncoded Stealth}
\label{sec:uncoded_stealth}
Let $\Pxn$ and $\Pxdn$ 
denote the marginals of input distributions satisfying \eqref{eq:power-constraint} and \eqref{eq:power-constraint-o}, respectively. Let $\Pzn$ and $\Pzdn$ be the corresponding marginals of the distributions $\Pzcw$ and $\Pzdprod$, respectively, which Warren observes at the output of his BSC$(q)$ from Alice.
To prevent Warren from detecting the communication with Bob, Alice must ensure that
\begin{align}
    \id{P_{Z^n}}{\Pzd^n} \leq \delta 
    \label{eq:stealth_coded}
\end{align}
for a small constant $\delta>0$.

Following \cite{wang2016fundamental}, we first consider an uncoded stealth scenario, where $Z^n$ is i.i.d., i.e., $\Pzcw = \Pzprod$.
The stealth constraint \eqref{eq:stealth_coded} is then
\begin{align}
n\id{\Pzn}{\Pzdn} \leq \delta \,.
\label{eq:stealth_uncoded}
\end{align}

Let $\muvec_n = \begin{bmatrix}\frac{a\na}{n} & \frac{b\nb}{n}\end{bmatrix}^T$. We write $\id{\Pzn}{\Pzdn}$ as a function of $\muvec_n$ and use
\begin{align}
\id{\Pzn}{\Pzdn} = \frac{1}{2} \frac{(\qbar-q)^2}{q\qbar} \left(\frac{a\na}{n}-\frac{b\nb}{n}\right)^2 + o(\norm{\muvec_n}^2).
\label{eq:taylor_uncoded_id_Q}
\end{align}
To prove \eqref{eq:taylor_uncoded_id_Q}, note that the second-order Taylor approximation for a scalar function $g:\:\cX^n\to\setreal$ around a point $\vecx_0$ is
\begin{align}
    g(\vecx) &= g(\vecx_0) +  \left.\nabla g(\vecx)^T\right\vert_{\vecx=\vecx_0} (\vecx-\vecx_0) \nonumber\\
    &\quad+  \frac{1}{2}(\vecx-\vecx_0)^T \left.\nabla^2 g(\vecx)\right\vert_{\vecx=\vecx_0} (\vecx-\vecx_0) + o(\norm{\vecx-\vecx_0}^2)
    \label{eq:Taylor_general_vector}
\end{align}
where $\nabla g$ and $\nabla^2 g$ denote the gradient and the Hessian matrix of $g$, respectively. We further have
\begin{align}
&\left.\id{\Pzn}{\Pzdn}\right\vert_{\muvec_n=\mathbf{0}} = 0 \label{eq:taylor_uncoded_id_d0}\\
&\left.\nabla\id{\Pzn}{\Pzdn}\right\vert_{\muvec_n=\mathbf{0}} = \mathbf{0} \label{eq:taylor_uncoded_id_d1}\\
&\left.\nabla^2\id{\Pzn}{\Pzdn}\right\vert_{\muvec_n=\mathbf{0}} = \begin{bmatrix}
\frac{(\qbar-q)^2}{q\qbar} & -\frac{(\qbar-q)^2}{q\qbar} \\ -\frac{(\qbar-q)^2}{q\qbar} & \frac{(\qbar-q)^2}{q\qbar} 
\end{bmatrix} \label{eq:taylor_uncoded_id_d2}\,.
\end{align}
Inserting \eqref{eq:taylor_uncoded_id_d0}--\eqref{eq:taylor_uncoded_id_d2} into \eqref{eq:Taylor_general_vector} gives \eqref{eq:taylor_uncoded_id_Q}.

From \eqref{eq:taylor_uncoded_id_Q}, the bound \eqref{eq:stealth_uncoded} is fulfilled for sufficiently large $n$ if
\begin{align}
\abs{a\na-b\nb} \leq k\sqrt{n} \qquad\text{with}\qquad k = \frac{\sqrt{2q\qbar}}{\qbar-q}\sqrt{\delta} .
\label{eq:stealth_uncoded_Q_2}
\end{align}
Alice can thus determine achievable values of $(a,\alpha)$ if Warren expects her to send with total transmit energy $b\nb$. Alternatively, Alice can determine how much energy to invest for obfuscation to keep Warren confused when she transmits information. 

A trivial but intuitive choice is $a\na = b\nb$. In this case, $\Pzn=\Pzdn$ and $\id{\Pzn}{\Pzdn} = 0$. Moreover, if $\beta>\frac{1}{2}$ (or $\alpha>\frac{1}{2}$), this is the only choice for which \eqref{eq:stealth_uncoded_Q_2} holds for all $n$. If we consider a fixed number of channel uses $n$, however, we can choose any values of $a\na$ and $b\nb$ satisfying \eqref{eq:stealth_uncoded_Q_2}.

\begin{figure}
	\centering
	\input{./fig2.tex}
	\caption{Information exponent $\alpha$ vs.\ obfuscation exponent $\beta$.}
	\label{fig:achievable_ab_kl0_uncoded}
\end{figure}
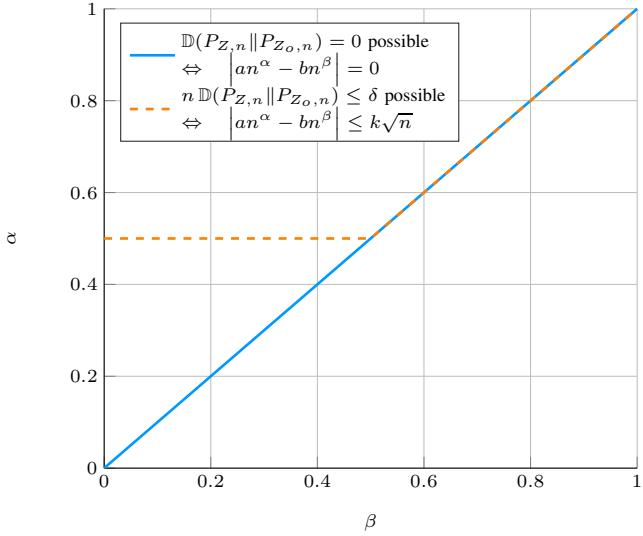

For large $n$, the \ac{LHS} of \eqref{eq:stealth_uncoded_Q_2} is dominated by the maximum exponent $\max(\alpha,\beta)$. If $\beta\leq\frac{1}{2}$, Alice could choose $\alpha=\frac{1}{2}$ and still satisfy the stealth constraint \eqref{eq:stealth_uncoded_Q_2}. Fig.~\ref{fig:achievable_ab_kl0_uncoded} summarizes the achievable information exponents $\alpha$ as a function of the obfuscation exponent $\beta$.

Consider now the covert communication scenario where $\frac{b\nb}{n}=0$. The \ac{LHS} of \eqref{eq:stealth_uncoded_Q_2} simplifies to
\begin{equation}
a\na \leq k\sqrt{n}
\end{equation}
and allows Alice to set $\alpha=\frac{1}{2}$ and $a=k$ for any $n$. 
We recover the square root law for covert communication with maximum codebook scaling constant (see \eqref{eq:Ra_max})
\begin{equation}
R_{\frac{1}{2},\text{max}} = k (1-2p) \log\frac{\pbar}{p} = \frac{\sqrt{2q\qbar}\cdot(1-2p)}{1-2q}\sqrt{\delta}\log\frac{\pbar}{p} \label{eq:Ralpha_05} .
\end{equation}

Fig.~\ref{fig:achievable_ab_kl0_uncoded} might give the impression that if one can transmit with $\alpha=\frac{1}{2}$ even if $\frac{b\nb}{n}=0$, then spending energy on obfuscation does not help in transmitting more information unless $\beta>0.5$. However, consider the case $\beta=\frac{1}{2}$. As before, Alice can choose the information exponent also to be $\alpha=\frac{1}{2}$. The \ac{LHS} of \eqref{eq:stealth_uncoded_Q_2} now reduces to $\abs{a-b} \leq k$. This allows Alice to choose
\begin{equation}
a = k + b
\end{equation}
which translates into an increased maximum square root scaling constant $R_{\frac{1}{2},\text{max}}$ compared with \eqref{eq:Ralpha_05}.

\subsection{Coded Stealth}
\label{sec:coded_stealth}
Consider the following random coding experiment. Alice generates $MK$ codewords $X^n(w,v)$, $w=1,\ldots,M$, $v=1,\ldots,K$, where the codeword symbols are choosen i.i.d. according to $\Pxn$. Let $\tcC = \lbrace X^n(1,1),\ldots,X^n(M,K) \rbrace$ be the random codebook and let all codewords be equiprobable. Further, Alice and Bob share a secret key $\tilde{v}$ drawn uniformly from $\{1,\ldots,K\}$. Let $\tcC_{\tilde{v}} = \lbrace X^n(1,\tilde{v}),\ldots,X^n(M,\tilde{v}) \rbrace$ be the corresponding subcodebook.

\emph{Alice:} Given a message $w$ and the key $\tilde{v}$, Alice transmits the codeword $x^n(w,\tilde{v})$ from the subcodebook $\tcC_{\tilde{v}}$.

\emph{Bob:} Bob observes the output $y^n$ of his BSC$(p)$ from Alice. As he knows that Alice used the subcodebook $\tcC_{\tilde{v}}$, he finds his \ac{ML} estimate as
\begin{align}
    \hat{w}=\argmax_{w'\in\{1,\ldots,M\}} P_{Y^n\given X^n}(y^n\given x^n(w',\tilde{v})) \,.
\end{align}

\emph{Warren:} Warren observes the output $z^n$ of his BSC$(q)$. To detect whether Alice was transmitting information to Bob or not, he runs a binary hypothesis test. As he does not know the secret key $\tilde{v}$, he must test against the entire codebook $\tcC$.

\emph{Reliability:} Both Alice and Bob know that subcodebook $\tcC_{\tilde{v}}$ was used. As the $nM$ symbols of $\tcC_{\tilde{v}}$ are sampled from $\Pxn$, we can apply the modified error exponents from Sec.~\ref{sec:mod_error_exponents}. According to \eqref{eq:Ra_max}, the probability of decoding error can be made small as long as
\begin{align}
    \Ra < a(1-2p)\log\frac{\pbar}{p} \,.
    \label{eq:Ra_reliability}
\end{align}

\emph{Stealth:} Warren observes $Z^n$ with the distribution
\begin{equation}
\PzcwtcC(z^n\given \tcC) = \sum_{w=1}^{MK} \frac{1}{MK} \Pzx^n(z^n\given X^n(w,v)) \,.
\end{equation}
To keep Warren confused, Alice must therefore ensure that
\begin{equation}
\expop\left[ \id{\PzcwtcC}{\Pzdprod} \right] \leq \theta
\label{eq:stealth_codebook}
\end{equation}
for a small constant $\theta>0$. 

Let $\Ra^{MK} = (\log M + \log K)/\na$ be the scaling constant of the code $\tcC$. The stealth constraint \eqref{eq:stealth_codebook} is satisfied by choosing
\begin{equation}
\Ra^{MK} > a(1-2q) \log\frac{\qbar}{q}
\label{eq:Ra_stealth}
\end{equation}
where $a$ satisfies the uncoded stealth constraint \eqref{eq:stealth_uncoded_Q_2} for an appropriately small constant $\delta>0$.
\begin{IEEEproof}
We split $\expop\left[ \id{\PzcwtcC}{\Pzdprod} \right]$ into three parts:
\begin{align}
&\expop\left[ \id{\PzcwtcC}{\Pzdprod} \right] = \underbrace{\expop\left[ \id{\PzcwtcC}{\Pzprod} \right]}_{(a)} + \underbrace{\id{\Pzprod}{\Pzdprod}}_{(b)} \nonumber\\ 
&+ \underbrace{\expop\left[ \left(\sum_{z_n}\PzcwtcC(z^n\given\tcC) - \Pzprod(z^n)\right) \log\frac{\Pzprod(z^n)}{\Pzdprod(z^n)} \right]}_{(c)} \,. \label{eq:stealth_proof_decomposition}
\end{align}
Bounding ($a$) is a standard \emph{resolvability} problem \cite{han1993approximation}, \cite{bloch2016covert}. We follow the proof technique from \cite[Lemma~2]{hayashi2006general}, \cite[Sec.~III-B]{hou2013informational}, \cite[Sec.~5.2.2, Lemma~5.3]{hou2014diss} that develops resolvability exponents, and we adapt it to our \ac{VP} transmission setting. Consider $-\frac{1}{2}\leq\rho\leq 0$. Since the modified error exponent $\Eoha$ as in \eqref{eq:E0ha_definition} and \eqref{eq:E0ha_simplified} becomes negative for these values of $\rho$, we define the modified resolvability exponents as
\begin{align}
& \Eorha(\rho,\Pxn) = -\Eoha(\rho,\Pxn) \label{eq:E0rha_definition}\\
& \Erha(\Ra,\Pxn) = \inf_{-\frac{1}{2}\leq\rho\leq 0} \left(\Eoha(\rho,\Pxn) + \rho\Ra^{MK} \right) \,.
\label{eq:Erha_definition}
\end{align}
The analysis in Sec.~\ref{sec:error_exponent} also holds for \eqref{eq:E0rha_definition} and \eqref{eq:Erha_definition}, and we obtain
\begin{align}\begin{cases}
\Erha(\Ra^{MK},\Pxn) <0 \quad \text{if } \Ra^{MK} > a(1-2q) \log\frac{\qbar}{q} \\
\Erha(\Ra^{MK},\Pxn) =0 \quad \text{if } \Ra^{MK} \leq a(1-2q) \log\frac{\qbar}{q} \,.\end{cases} \label{eq:Erha_cases}
\end{align}
Next, following \cite[Sec.~III]{hayashi2006general}, the average divergence $\expop\left[ \id{\PzcwtcC}{\Pzprod} \right]$ is the mutual information $I(\tcC;Z^n)$ of the codebook $\tcC$ and the channel output $Z^n$. We therefore define
\begin{align}
\Eohan(\rho,\Pxcw) &= \log \sum_{z^n} \left\{ \expop\left[\PzcwtcC(z^n\given\tcC) \right]^{\frac{1}{1+\rho}} \right\}^{1+\rho} 
\end{align}
which has the following properties \cite[Lemma~2]{hayashi2006general}:
\begin{align}
&\Eohan(0,\Pxcw) = \Eorha(0,\Pxn) = 0 \label{eq:E0rhan_prop1}\\
&\left.\frac{\partial \Eohan(\rho,\Pxn)}{\partial\rho}\right\vert_{\rho=0} = -\expop\left[ \id{\PzcwtcC}{\Pzprod} \right]\label{eq:E0rhan_prop2}\\
&\left.\frac{\partial^2 \Eohan(\rho,\Pxn)}{\partial\rho^2}\right\vert_{\rho=0} \geq 0 \label{eq:E0rhan_prop3}\,.
\end{align}

A slight modification of the proof of \cite[Lemma~2]{hou2013informational}, \cite[Lemma~5.3]{hou2014diss} where we replace the codebook size by $\log |\tcC| = \na\Ra^{MK}$ in \cite[Eq.~(46)]{hou2013informational} yields
\begin{align}
\Eohan(0,\Pxcw) &\leq \log\left(1+e^{\na\Erha(\Ra^{MK} ,\Pxn)}\right)\nonumber\\
&\leq e^{\na\Erha(\Ra^{MK} ,\Pxn)} . \label{eq:lemma53_mod}
\end{align}
By combining \eqref{eq:E0rhan_prop1}--\eqref{eq:E0rhan_prop3}, we obtain
\begin{align}
\rho\cdot\left(-\expop\left[ \id{\PzcwtcC}{\Pzprod} \right]\right) &\leq \Eohan(\rho,\Pxcw)
\end{align}
for $-\frac{1}{2}\leq\rho\leq 0$, and thus
\begin{align}
\expop\left[ \id{\PzcwtcC}{\Pzprod} \right] &\leq \frac{\Eohan(0,\Pxcw)}{-\rho} \nonumber\\
&\leq \frac{e^{\na\Erha(\Ra^{MK} ,\Pxn)}}{-\rho} \label{eq:final_bound_a}
\end{align}
where we used \eqref{eq:lemma53_mod} in the last step. By \eqref{eq:final_bound_a} and \eqref{eq:Erha_cases}, we see that the term ($a$) in \eqref{eq:stealth_proof_decomposition} goes to zero for $n\to\infty$ if $\Ra^{MK} >  a(1-2q) \log\frac{\qbar}{q}$.

To bound the term ($b$) in \eqref{eq:stealth_proof_decomposition} we note that $\id{\Pzprod}{\Pzdprod} = n \id{\Pzn}{\Pzdn}$. We can therefore reuse our results for uncoded stealth and must only ensure that we satisfy \eqref{eq:stealth_uncoded_Q_2} for a small enough constant $\delta$, $0<\delta<\theta$.

Finally, we rewrite term ($c$) in \eqref{eq:stealth_proof_decomposition} as follows: 
\begin{align}
&\abs{\expop\left[ \left(\sum_{z_n}\PzcwtcC(z^n\given\tcC) - \Pzprod(z^n)\right) \log\frac{\Pzprod(z^n)}{\Pzdprod(z^n)} \right]} \nonumber\\
&\leq 2n \expop\left[ \vd{\PzcwtcC}{\Pzprod}\right] \log\frac{1}{\nu_d}  \label{eq:bound_c_1}\\
&\leq 2n \sqrt{\frac{1}{2}\expop\left[ \id{\PzcwtcC}{\Pzprod}\right]} \log\frac{1}{\nu_d} \label{eq:bound_c_2}\\
&\leq \sqrt{2}n \cdot \frac{e^{\frac{1}{2}\na\Erha(\Ra^{MK} ,\Pxn)}}{\sqrt{-\rho}} \log\frac{1}{\nu_d} \label{eq:final_bound_c}
\end{align}
where we used \eqref{eq:vd_definition} and $\nu_d = \min_{z}\Pzdn(z)$ in \eqref{eq:bound_c_1}, where \eqref{eq:bound_c_2} follows by Pinsker's inequality \eqref{eq:pinsker_ineq} and Jensen's inequality, and where we reused the bound \eqref{eq:final_bound_a} in \eqref{eq:final_bound_c}. Again, the \ac{RHS} of \eqref{eq:final_bound_c} goes to zero as $n\to\infty$ if $\Ra^{MK} > a(1-2q) \log\frac{\qbar}{q}$.
\end{IEEEproof} 

Summarizing \eqref{eq:Ra_reliability} and \eqref{eq:Ra_stealth}, for small positive $\xi$ we can bound
\begin{align}
    &\log M \leq {\na} (1-\xi) a(1-2p)\log\frac{\pbar}{p} \label{eq:stealth_max_M}\\
    &\log K \geq \nonumber\\
    &{\na} \left[ (1+\xi) a(1-2q) \log\frac{\qbar}{q} - (1-\xi) a(1-2p)\log\frac{\pbar}{p}\right]^+ \label{eq:stealth_max_K}
\end{align}
where $[x]^+ = \max(x,0)$ and where $\alpha$, $a$ satisfy \eqref{eq:stealth_uncoded_Q_2} for specified $\beta$, $b$.

\section{Discussion} \label{sec:conclusion}
We  compare our results to bounds derived in \cite{wang2016fundamental,bloch2016covert,che2014reliable}.
The work in \cite{wang2016fundamental} considers covert communication where Warren and Bob both observe channel outputs from a BSC$(q)$. Moreover, the channel outputs are i.i.d. also when Alice transmits information to Bob, which is equivalent to our uncoded stealth scenario from Sec.~\ref{sec:uncoded_stealth} with $\alpha=\frac{1}{2}$ and $\frac{b\nb}{n}=0$. The maximum scaling constant \eqref{eq:Ralpha_05} reduces to $R_{\frac{1}{2},\text{max}} = \sqrt{2q\qbar}\sqrt{\delta}\log\frac{\qbar}{q}$, which is the same value one would obtain from \cite[Thm.~2]{wang2016fundamental} for \acp{BSC}.

Similarly, we compare our coded results \eqref{eq:stealth_max_M} and \eqref{eq:stealth_max_K} to the bounds in Corollary 2 of Thm.~2 in \cite{bloch2016covert} for the covert communication scenario. From \eqref{eq:stealth_uncoded_Q_2} we know that Alice can choose at most $\alpha=\frac{1}{2}$ and $a=k$. Further, $k$ can be alternatively expressed as 
\begin{align}
    k = \sqrt{\frac{2}{\chsq{\Pzx(\cdot\given 1)}{\Pzx(\cdot\given 0)}}}\cdot\sqrt{\delta}
\end{align}
where $\delta<\expop\left[ \id{\PzcwtcC}{\Pzdprod} \right]$. Our bounds then match the ones from \cite[Corollary~2]{bloch2016covert} when evaluated for \acp{BSC}. 

Our results also apply to covert communication without a secret key by choosing $\log K = 0$. The codebook scaling constant $\Ra$ is then upper and lower bounded by the \acp{RHS} of \eqref{eq:Ra_reliability} and \eqref{eq:Ra_stealth}, respectively. These are exactly the same bounds reported in \cite[Thm.~1]{che2014reliable} for \acp{BSC} and variational distance as the stealth measure, where the authors assumed that Bob's channel from Alice must be better than Warren's. Note that without a secret key, one must have $p<q$ to satisfy the bounds. Moreover, \eqref{eq:stealth_max_K} implies that the key size can be zero if $p<q$.

We conclude with two remarks. %
First, due to space constraints we presented only the achievability proof and left the converse proof for a future document. Second, we have studied \acp{BSC} only; extensions to general \acp{DMC} will be treated in a future document.

\section*{Acknowledgment}
This work was supported by the German Research Foundation (DFG) under Grant KR~3517/9-1.

\IEEEtriggeratref{7}
\bibliographystyle{IEEEtran}
\bibliography{IEEEabrv,confs-jrnls,references}

\end{document}

%% file: preamble.tex
\usepackage{amssymb}
\usepackage{xspace}
\usepackage{bbm}
\usepackage{booktabs}
\usepackage{graphicx}
\usepackage{tikz}
\usepackage{pgfplots}
\usepackage{trsym}
\usepackage{siunitx}
\usepackage{csquotes}
\usepackage{mathtools}

\newcommand{\setreal}{\ensuremath{\mathbf{R}}\xspace}

\DeclareMathOperator{\expop}{\mathbb{E}}

\DeclareMathOperator{\idop}{\mathbb{D}}
\DeclareMathOperator{\vdop}{\mathbb{V}}

\DeclareMathOperator*{\argmax}{argmax}

\newcommand{\vecx}{\boldsymbol{x}}

\newcommand{\cC}{\mathcal{C}}

\newcommand{\cX}{\mathcal{X}}

\newcommand{\bpm}{\begin{pmatrix}}
\newcommand{\epm}{\end{pmatrix}}
\newcommand{\bbm}{\begin{bmatrix}}
\newcommand{\ebm}{\end{bmatrix}}

\definecolor{TUMBeamerYellow}    {rgb}{1.00,0.71,0.00}  %
\definecolor{TUMBeamerOrange}    {rgb}{1.00,0.50,0.00}  %
\definecolor{TUMBeamerRed}       {rgb}{0.90,0.20,0.09}  %
\definecolor{TUMBeamerDarkRed}   {rgb}{0.79,0.13,0.25}  %
\definecolor{TUMBeamerBlue}      {rgb}{0.00,0.60,1.00}  %
\definecolor{TUMBeamerLightBlue} {rgb}{0.25,0.75,1.00}  %
\definecolor{TUMBeamerGreen}     {rgb}{0.57,0.67,0.42}  %
\definecolor{TUMBeamerLightGreen}{rgb}{0.71,0.79,0.51}  %

\newcommand{\equivalent}{\triangleq}
\newcommand{\given}{\:\!\vert\:\!}
\providecommand{\abs}[1]{\ensuremath{\left\lvert#1\right\rvert}}
\providecommand{\norm}[1]{\ensuremath{\lVert#1\rVert}}
\DeclarePairedDelimiterX{\infdivx}[2]{(}{)}{#1\delimsize\|#2}
\newcommand{\id}{\idop\infdivx}
\DeclarePairedDelimiterX{\vardistx}[2]{(}{)}{#1\delimsize\|#2}
\newcommand{\vd}{\vdop\vardistx}
\DeclarePairedDelimiterX{\ch2distx}[2]{(}{)}{#1\delimsize\|#2}
\newcommand{\chsq}{\chi_2\ch2distx}

%% file: fig1.tex
\pagestyle{empty}

\tikzstyle{init} = [minimum size=2em, align=center]
\tikzstyle{blck}=[rectangle, draw, minimum size=1pt, align=center]
\tikzstyle{blck2}=[rectangle, rounded corners, draw, minimum size=1pt, align=center]

\begin{tikzpicture}[auto,>=latex']

\node[draw, minimum width=1.5cm, minimum height=1cm] (alice) at (0,0) {Alice};

\node[draw, anchor=west] (prob) at (3,0) {$P^n_{Y|X}$};
\draw[->] (alice) -- (prob) node[pos=0.5, above] (X) {$X^n$};

\node[draw, anchor=west, minimum width=1.5cm, minimum height=1cm] (bob) at (6,0) {Bob};
\draw[->] (prob) -- (bob) node[pos=0.5, above] {$Y^n$};

\node[draw, anchor=west] (prob2) at (3, -1.5) {$P^n_{Z|X}$};

\draw[->] (X)  -- (X |- prob2) -- (prob2);

\node[draw, anchor=west, minimum width=1.5cm, minimum height=1cm] (div) at (6, -1.5) {Warren};
\draw[->] (prob2) -- (div) node[pos=0.5, above] {$Z^n$};

\end{tikzpicture}

%% file: fig2.tex
\begin{tikzpicture}
\scriptsize
\begin{axis}[%
width=0.8*\columnwidth,
scale only axis,
xmin=0,
xmax=1,
xlabel={$\beta$},
xmajorgrids,
ymin=0,
ymax=1,
ylabel={$\alpha$},
ymajorgrids,
axis x line*=bottom,
axis y line*=left,
legend style={at={(0.03,0.97)},anchor=north west,legend cell align=left,align=left,draw=white!15!black}
]
\addplot [color=TUMBeamerBlue,solid,line width=1.0pt]
  table[row sep=crcr]{%
  0 0\\
  1 1\\
};
\addlegendentry{$\id{\Pzn}{\Pzdn} = 0$ possible\\$\Leftrightarrow \quad \abs{a\na-b\nb} =0$};
\addplot [color=TUMBeamerOrange,dashed,line width=1.0pt]
  table[row sep=crcr]{%
  0 0.5\\
  0.5 0.5\\
  1 1\\
};
\addlegendentry{$n\id{\Pzn}{\Pzdn} \leq\delta$ possible\\$\Leftrightarrow \quad \abs{a\na-b\nb} \leq k\sqrt{n}$};

\end{axis}
\end{tikzpicture}%